\documentclass{emulateapj}

\slugcomment{To Appear in the Astrophysical Journal Letters}

\shorttitle{Transition of Clustering for LBGs}
\shortauthors{Ouchi et al.}

\begin{document}


\title{Definitive Identification of the Transition between \\
Small- to Large-Scale Clustering for Lyman Break Galaxies\altaffilmark{1}}


\author{Masami Ouchi        \altaffilmark{2,3},
        Takashi Hamana      \altaffilmark{4},
	Kazuhiro Shimasaku  \altaffilmark{5},
	Toru Yamada         \altaffilmark{4},\\
        Masayuki Akiyama    \altaffilmark{6},
        Nobunari Kashikawa  \altaffilmark{4},
        Makiko Yoshida      \altaffilmark{5},
	Kentaro Aoki        \altaffilmark{6},\\
	Masanori Iye        \altaffilmark{4},
	Tomoki Saito        \altaffilmark{5},
	Toshiyuki Sasaki    \altaffilmark{6},\\
	Chris Simpson       \altaffilmark{7}, and
	Michitoshi Yoshida  \altaffilmark{8}
        }

%\email{ouchi@stsci.edu}

%\author{S. Djorgovski\altaffilmark{1,2,3} and Ivan R. King\altaffilmark{1}}
%\affil{Astronomy Department, University of California,
%    Berkeley, CA 94720}
%
%\author{C. D. Biemesderfer\altaffilmark{4,5}}
%\affil{National Optical Astronomy Observatories, Tucson, AZ 85719}
%\email{aastex-help@aas.org}
%
%\and
%
%\author{R. J. Hanisch\altaffilmark{5}}
%\affil{Space Telescope Science Institute, Baltimore, MD 21218}


\altaffiltext{1}{Based on data collected at 
        Subaru Telescope, which is operated by 
        the National Astronomical Observatory of Japan.}
\altaffiltext{2}{Space Telescope Science Institute,
        3700 San Martin Drive, Baltimore, MD 21218, USA; ouchi@stsci.edu.}
\altaffiltext{3}{Hubble Fellow}
\altaffiltext{4}{National Astronomical Observatory, 
        Tokyo 181-8588, Japan}
\altaffiltext{5}{Department of Astronomy, School of Science,
        University of Tokyo, Tokyo 113-0033, Japan}
\altaffiltext{6}{Subaru Telescope, National Astronomical Observatory, 
        650 N.A'ohoku Place, Hilo, HI 96720, USA}
\altaffiltext{7}{Department of Physics, University of Durham,
        South Road, Durham DH1 3LE, UK}
\altaffiltext{8}{Okayama Astrophysical Observatory,
    National Astronomical Observatory, Kamogata, Okayama 719-0232, Japan}

%\altaffiltext{1}{Visiting Astronomer, Cerro Tololo Inter-American Observatory.
%CTIO is operated by AURA, Inc.\ under contract to the National Science
%\altaffiltext{2}{Society of Fellows, Harvard University.}
%\altaffiltext{3}{present address: Center for Astrophysics,
%    60 Garden Street, Cambridge, MA 02138}
%\altaffiltext{4}{Visiting Programmer, Space Telescope Science Institute}
%\altaffiltext{5}{Patron, Alonso's Bar and Grill}


%The abstract should concisely summarize the content and conclusions of 
%the paper. The abstract should be a single paragraph of not more than 
%250 words and should not contain reference citations. 

\begin{abstract}
We report angular correlation function (ACF) of
Lyman Break Galaxies (LBGs) with unprecedented statistical 
quality on the basis of 16,920 LBGs at $z=4$ 
detected in the 1 deg$^2$ sky of 
the Subaru/XMM-Newton Deep Field.
The ACF significantly departs from a power law,
and shows an excess on small scale.
Particularly, the ACF of LBGs with $i'<27.5$ have a clear break
between the small and large-scale regimes at the 
angular separation of $\simeq 7''$ whose projected
length corresponds to the virial radius of dark halos
with a mass of $10^{11-12}M_\odot$, indicating
multiple LBGs residing in a single dark halo.
Both on small ($2''<\theta<3''$) and large ($40''<\theta<400''$)
scales, clustering amplitudes monotonically increase
with luminosity for 
the magnitude range of $i'=24.5-27.5$,
and the small-scale clustering 
shows a stronger luminosity dependence than the large-scale clustering.
The small-scale bias reaches $b \simeq 10-50$, and
the outskirts of small-scale excess 
extend to a larger angular separation for brighter LBGs.
The ACF and number density
%of LBGs are reasonably reproduced by 
%a halo model
%in the framework of the halo occupation distribution
%of the cold dark matter model.
of LBGs can be explained by 
the cold dark matter model.
\end{abstract}


%% subject header.  Objects should be in the appropriate "individual"
%% headers (e.g. quasars: individual, stars: individual, etc.) with the
%% additional provision that the total number of headers, including each
%% individual object, not exceed six.  The \objectname{} macro, and its
%% alias \object{}, is used to mark each object.  The macro takes the object
%% name as its primary argument.  This name will appear in the paper
%% and serve as the link's anchor in the electronic edition if the name
%% is recognized by the data centers.  The macro also takes an optional
%% argument in parentheses in cases where the data center identification
%% differs from what is to be printed in the paper.


%globular clusters: general ---
%globular clusters: individual(\objectname{NGC 6397},
%\object{NGC 6624}, \objectname[M 15]{NGC 7078},
%\object[Cl 1938-341]{Terzan 8})}


\section{Introduction}
\label{sec:introduction}
%Recently, there has been great progress in the observations 
%of large-scale structures at high redshifts.
%Strong clustering has been found in the
Recent observational studies have been found
strong clustering in
two-point angular correlation function (ACF) of 
Lyman break galaxies (LBGs) at $z=3-5$,
(e.g. \citealt{giavalisco2001,ouchi2001,foucaud2003,
adelberger2003,ouchi2004b,hildebrandt2004,allen2005}),
red galaxies at $z=3$ \citep{daddi2003},
and Ly$\alpha$ emitters (LAEs; \citealt{ouchi2003,shimasaku2004})
at $z=5$. Even at $z=6$, there is a piece of 
evidence for filamentary large (100 Mpc)-scale 
structures of LAEs \citep{ouchi2005}.
%whose rms fluctuation
%is comparable to those of present-day galaxies \citep{ouchi2005}.
%The strong clustering indicates 
The distribution
of high-$z$ galaxies is fairly inhomogeneous
and highly biased 
against
matter distribution predicted by 
the cold dark matter (CDM) model.
%The estimated bias is in the range of $b\simeq 2-8$, 
%
%% revision cut start
The estimated bias is $b\simeq 2-8$, 
%
%% revision cut end
%
depending on luminosity/type and redshift of galaxies.
However, the shape of the ACF for high-$z$ galaxies is not well
%
%constrained; 
%i.e., \citet{ouchi2001} reported the significant
%excess of ACFs at $\theta<5''$,
%while \citet{porciani2002} 
%found that the ACF 
%at $10''\lesssim\theta\lesssim30''$ becomes small.
%
%% revision other2 start
%
%{\bf
constrained. \citet{ouchi2001}
report a $3\sigma$ excess of the ACF at $\theta<5''$ for $z\sim4$ LBGs,
while \citet{porciani2002} found a possible deficit
of the ACF at $10''\lesssim\theta\lesssim30''$ for bright LBGs 
which they interpret as the halo exclusion effect on hosting halos
with a mass of $10^{12} M_\odot$.
%}
%
%% revision other2 end
%

In the local universe, the correlation function 
shows a departure from a power law 
%(e.g. \citealt{connolly2002,hawkins2003,zehavi2004}).
%
%% revision cut start
(e.g. \citealt{hawkins2003,zehavi2004}).
%
%% revision cut end
%
The departure is reproduced in the framework of 
the halo occupation distribution (HOD) and the related
halo models in the CDM cosmology
\citep{vandenbosch2003,magliocchetti2003,zehavi2004,
benson2001,berlind2003}, and is explained by
two sources contributing to the correlation function;
one for galaxy pairs residing in the same halo (1-halo term)
and the other for galaxies hosted by different halos (2-halo term;
see, e.g., \citealt{zehavi2004}).
The HOD has been also applied to clustering of galaxies at high-$z$
\citep{bullock2002,moustakas2002,hamana2004}.
However, parameters of the models have not been constrained 
with similar accuracy as at low-$z$
%, because the sample size 
%and surveyed area are still relatively small,
%i.e., $0.01-0.1$deg$^2$ and $100-2000$, respectively.
%
%% revision other1 start
%
%{\bf
because of the small sample ($100-2000$ galaxies)
and surveyed area ($0.01-0.1$deg$^2$).
%i.e., $100-2000$ and $0.01-0.1$deg$^2$, respectively.
%}
%
%% revision other1 end
%

In this paper, we present ACF of $z=4$ LBGs
with unprecedented statistical quality,
%on the basis of large high-$z$ galaxy sample (16,920 LBGs)
%obtained by the wide-field (1 deg$^2$) survey in
%Subaru/XMM-Newton Deep Field (SXDF; \citealt{sekiguchi2004}).
on the basis of 16,920 LBGs obtained in the 1 deg$^2$ sky of
Subaru/XMM-Newton Deep Field (SXDF; \citealt{sekiguchi2004}).
Throughout this paper, magnitudes are in the AB system,
and we adopt $H_0=70 h_{70} $km s$^{-1}$ Mpc$^{-1}$ and
$[\Omega_m,\Omega_\Lambda,n,\sigma_8]=[0.3,0.7,1.0,0.9]$.
%To facilitate comparison with previous results, we express 
%$r_0$ using $h_{100}$, where $h_{100}$ is 
%the Hubble constant in units of 100 km s$^{-1}$ Mpc$^{-1}$.
To facilitate comparison with previous results,
we express $r_0$ using $h_{100}$, the Hubble constant 
in units of 100 km s$^{-1}$ Mpc$^{-1}$.

\begin{figure}
%\epsscale{0.7}
\epsscale{1.15}
%\epsscale{1.05}
%\plotone{image/dist_BRiLBG.eps}
%\plotone{f1.eps}
\plotone{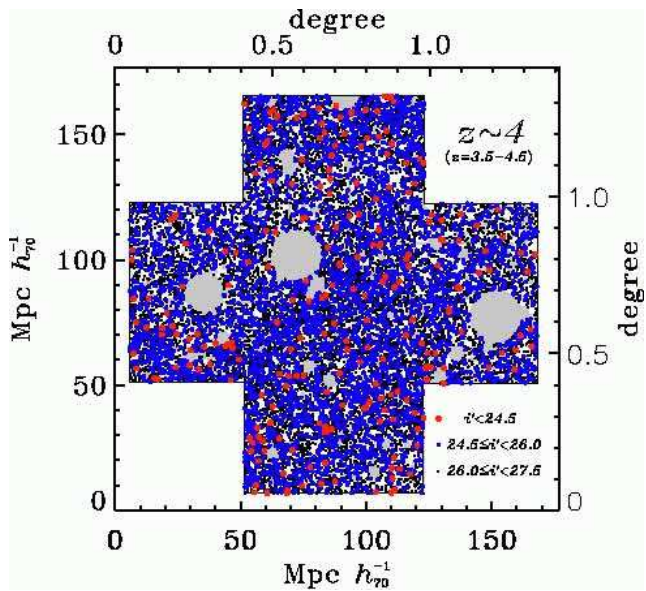}
\caption{
The distribution of LBGs at $z=4.0\pm 0.5$
in the SXDF.
The red, blue, and black points denote the positions
of the LBGs with $i'<24.5$ (bright), $24.5\le i'<26.0$ (intermediate),
and $26.0\le i'<27.5$ (faint), respectively. The gray areas
present masked regions 
%contaminated with bright stars 
%and CCD blooming 
where we did not use for our analysis.
The scale on the map is marked in both degrees and 
%the projected distance in comoving megaparsecs at $z=4.0$. 
(comoving) megaparsecs for projected distance at $z=4.0$.
{\bf This figure is degraded. This paper with the original figure
can be downloaded from
$http://www-int.stsci.edu/\sim ouchi/work/astroph/sxds\_z4LBG/ouchi\_highres.pdf$ }
\label{fig:dist_BRiLBG}}
\end{figure}

%\clearpage

\section{Data and Sample}
\label{sec:data_and_sample}
We carried out deep optical broad-band imaging
%with Subaru/Suprime-Cam \citep{miyazaki2002} in the 1 deg$^2$
%
%% revision cut start
with Subaru/Suprime-Cam in the 1 deg$^2$
%
%% revision cut end
%
sky of the SXDF. 
%Our broad-band ($B$, $V$, $R$, $i'$, and $z'$) images reach
%
%% revision cut start
%
Our broad-band images reach
%
%% revision cut end
%
$B\simeq28.3$, $V\simeq 27.3$,
$R\simeq27.6$, $i'\simeq27.5$, and $z'\simeq 26.5$
with a $2''$-diameter circular aperture
at the $3\sigma$ level
(Furusawa et al. in preparation).
Typical seeing sizes (FWHM) of these images
are $0''.8$.
We use the $i'$-band selected source catalog 
of the SXDS $Ver1.0$
%
%% revision 2)-1 start
%
%{\bf
produced with SExtractor \citep{bertin1996}, 
%}
%
%% revision 2)-1 end
%
which is composed of 0.7 million objects
with $i'<27.5$.
%
%i<27.5 : 725,836 objects
%
We select LBGs at $z=4.0\pm 0.5$ on the basis of the color criteria of 
\citet{ouchi2004a}, i.e.,
$B-R>1.2$, $R-i'<0.7$, and $B-R>1.6(R-i')+1.9$, which were determined
with the results of spectroscopy and Monte-Carlo simulations. 
%Twenty thousand LBG candidates satisfy the color criteria.
We visually inspect all the candidates and mask
areas contaminated with halos of bright stars and CCD blooming.
%Our final catalog includes 16,920 LBGs in a 1.00 deg$^2$ area.
%Figure \ref{fig:dist_BRiLBG}
%shows the sky distribution of our LBGs. 
%We summarize our LBG samples
%in Table \ref{tab:results}.
%
%% revision cut start
%
Our final catalog includes 16,920 LBGs in a 1.00 deg$^2$ area 
(Table \ref{tab:results}).
Figure \ref{fig:dist_BRiLBG}
shows the sky distribution of our LBGs. 
%
%% revision cut end
%
Our spectroscopic follow-up observations show that
60 out of 63 identified candidates are real LBGs
at $z=3.5-4.5$; i.e.,
17 out of 17 and 43 out of 46 are LBGs
in the SXDF (Akiyama M. in preparation) and 
in the Subaru Deep Field, respectively,
where the latter LBG sample is made 
with the same color criteria as ours \citep{yoshida2005}.
Thus, the contamination rate of our LBG sample is estimated to be
$(63-60)/63 = 5$\%.

\begin{figure}
%\epsscale{0.85}
\epsscale{1.20}
%\plotone{image/acorr_BRiLBG_all.eps}
\plotone{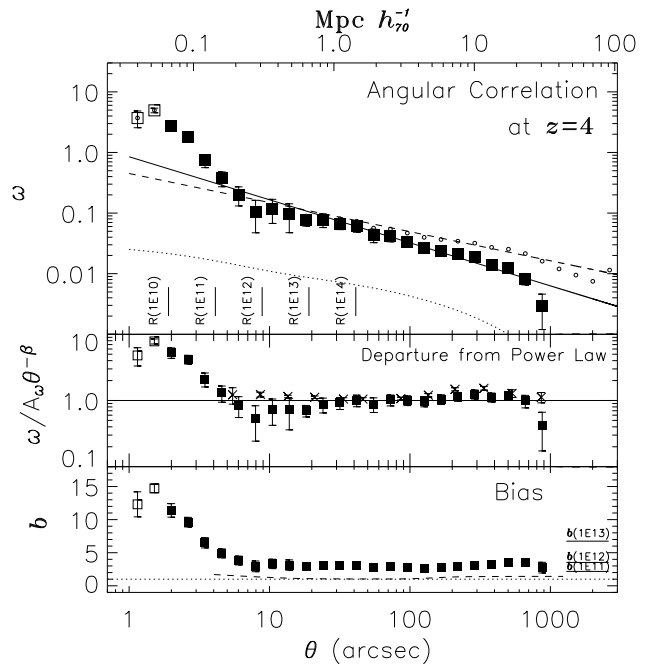}
\caption{
$Top:$ The ACF, $\omega(\theta)$, of 
LBGs. 
%The squares and error bars indicate
%the ACF and 1 $\sigma$ bootstrap errors
%of LBGs.
The filled and open squares indicate the ACF with
1 $\sigma$ bootstrap errors, while
the open squares mean for ACF on very small scale
which may include additional errors in source deblending and confusion.
The solid line is
the best-fit power law ($A_\omega \theta^{-\beta}$) 
for $2''-1000''$.
The open circles are the ACF with IC correction
under the assumption of the conventional power-law approximation,
and the dashed lines are the best-fit power law
for these open circles.
The dotted curve is the ACF of 
dark matter predicted by the non-linear model
of \citet{peacock1996}. The scale on the top axis denotes
the projected distance in comoving megaparsecs at $z=4.0$. 
The ticks labeled with R(1E10), R(1E11), R(1E12),
R(1E13), and R(1E14) correspond to the predicted virial radii of 
dark halos, $r_{200}$, with a mass
of $1\times 10^{10}$, $10^{11}$, 
$10^{12}$, $10^{13}$, and
$10^{14}$ $h_{70}^{-1}M_\odot$, respectively.
%, within which
%the mean enclosed density is 200 times the mean cosmic value
%\citep{mo2002}.
$Middle:$ The ratios of the ACF to the best-fit power law
for our LBGs (squares), together with those for local galaxies
(crosses; \citealt{zehavi2004}).
$Bottom:$ The galaxy-dark matter bias, $b$, of LBGs
as a function of separation. The dashed curve presents
bias of local galaxies \citep{zehavi2004}.
The ticks with $b$(1E11),
$b$(1E12), and $b$(1E13) show linear biases of dark halos with
a mass of $1\times 10^{11}$, 
$10^{12}$, and 
$10^{13}$ $h_{70}^{-1}M_\odot$, respectively,
predicted by the CDM model of \citet{sheth1999}.
\label{fig:acorr_BRiLBG_all}}
\end{figure}

%\clearpage

\section{Results and Discussion}
\label{sec:results_and_discussion}
%\subsection{Definitive Identification of the Transition between Small and Large-Scale Clustering}
%
%% revision cut start
%
\subsection{Definitive Identification of Clustering Transition}
%
%% revision cut end
%
\label{subsec:definitive_identification}
%We derive the ACF, $\omega(\theta)$, by the formula of
%\cite{landy1993}
%with random samples composed of 200,000 sources, 
%and estimate bootstrap errors \citep{ling1986}.
%
%% revision2-3 start
%
We derive the ACF, $\omega(\theta)$, by the formula of
\cite{landy1993} with random samples composed of 200,000 sources,
and estimate bootstrap errors \citep{ling1986}.
%Note that bootstrap errors are assumed to be
%independent of angular separations.
%
%% revision2-3 end
%
Since clustering properties of our 5\% contaminants
are not clear, we do not apply a correction
for contaminants with the assumption of random distribution
(c.f. \citealt{ouchi2004b}). However, 
this correction changes $\omega(\theta)$ and bias
only by $10$\% or less.
Figure \ref{fig:acorr_BRiLBG_all} presents
the ACF of LBGs (top panel), residuals of a power-law fit (middle panel),
and galaxy-dark matter bias (bottom panel) defined as
$
b(\theta)
 \equiv
  \sqrt{\omega(\theta)/\omega_{\rm dm}(\theta)}
$,
where $\omega_{\rm dm}(\theta)$ is the ACF predicted 
by the non-linear model of \cite{peacock1996}.
In the top and middle panels of Figure \ref{fig:acorr_BRiLBG_all}
the ACF of LBGs shows a significant excess on small scale, 
and indicates that a power law,
$A_\omega \theta^{-\beta}$, 
does not fit the data.
This is the definitive identification of the departure from a power law
for the ACF of LBGs at $z=4$. 
%It should be noted that we have visually inspected 
%
%we visually inspect
%all close-pairs of LBGs
%and 
%
%Moreover, a very recent study by \citet{lee2005}
%also finds a similar small-scale excess for 
%$z=4-5$ LBGs in a sample obtained by HST/ACS-GOODS.
%
%% revision 1) + 2)-2 start
%
%{\bf
%Although most of our LBGs are marginally resolved (FWHM $\simeq 1''$),
%the uncertainties in source deblending and photometry may account 
%for the small-scale excess of the ACF at less than $\simeq 2''$.
%However, the small-scale excess at $\gtrsim 2''$ cannot be explained by
%such uncertainties.
%In fact, a very recent study based on high-resolution ($\sim 0''.1$) HST images
%\citep{lee2005} 
%also finds a similar excess on $2''-10''$ for $z=4-5$ LBGs. 
%}
%
%% revision 1) + 2)-2 end
%
%
%% revision2-1 start
%
With a visual inspection,
we confirm that all close-pairs of LBGs are not false detections.
We also plot histogram of galaxy sizes for LBG pairs. 
We find that most of our LBGs have FWHM$\simeq 1''$ for pairs with any separations 
down to, at least, $\simeq 2''$, and that extended LBGs do 
not boost small-scale ACF by producing false pairs. Uncertainties in source 
deblending and photometry can hardly account for the small-scale 
excess at $\gtrsim 2''$. In fact,
a similar small-scale excess of ACF for $z=4-5$ LBGs is also found by 
a recent study on high-resolution ($\sim 0''.1$) HST images \citep{lee2005}.
%
%% revision2-2 end
%

Comparing our ACF with 
the one of dark matter, we find that the small-scale 
excess extends up to $\simeq 7''$, i.e. $0.24 h_{70}^{-1}$ Mpc,
which is comparable to virial radius, $r_{200}$,
of dark halos with a mass of $10^{11-12}M_\odot$
(see the ticks in the top panel of Figure \ref{fig:acorr_BRiLBG_all}),
where $r_{200}$ is a sphere of radius within which
the mean enclosed density is 200 times the mean cosmic value
\citep{mo2002}.
Interestingly, the large-scale average bias
at $40''< \theta < 400''$
is estimated to be $2.9\pm 0.2$
which is also comparable to linear bias of dark halos
with a mass of $10^{11-12}M_\odot$ ($b=2.2-3.5$)
predicted by the CDM (\citealt{sheth1999};
see the ticks in the bottom panel of Figure \ref{fig:acorr_BRiLBG_all}).
This coincidence of the dark-halo mass strongly supports
that typical $z=4$ LBGs reside in dark halos with a mass of 
$10^{11-12}M_\odot$. Moreover, these pieces of evidence
suggest that multiple LBGs occupy a single dark halo.
%, which can be modeled with the HOD. 
The middle panel of Figure \ref{fig:acorr_BRiLBG_all} also plots
residuals of a power-law fit for local galaxies \citep{zehavi2004},
which is comparable to those of $z=4$ LBGs on large scale, but
significantly larger than LBGs on intermediate scale ($0.2-1.0$ Mpc)
corresponding to the radius of $10^{12-14} M_\odot$ dark halos.
According to the halo mass function \citep{sheth1999},
the ratio of galaxy-sized halos ($10^{10-12}M_\odot$) to
group/cluster-sized halos ($10^{12-14}M_\odot$) is about 10 times
higher in number density at $z=4$ than at $z=0$. 
This relative deficit of group/cluster-sized
halos at high-$z$ would 
be the cause of the clearer break
between small and large-scale ACFs at $z=4$
than $z=0$.
%
%% revision 1b) start
%
%{\bf
Although multiple occupation of LBGs
explains very consistently both the angular scale of transition and 
the amplitude of large-scale bias, 
there remains the possibility that
the small-scale excess is enhanced or produced by
brightening of pair galaxies due to interactions.
%}
%
%% revision 1b) end
%

%
%% revision other3 start
%
%The small-scale bias of LBGs shown in the bottom panel of
%Figure \ref{fig:acorr_BRiLBG_all}
%is particularly interesting.
%In the present-day universe, 
%late-type galaxies are antibiased on small scale,
%while early-type galaxies are positively biased
%\citep{vandenbosch2003}.
%This is often interpreted as due to
%the environmental effects of galaxy distribution 
%(e.g. \citealt{dressler1980}).
%Although LBGs are star-forming galaxies classified as
%the spectral type similar to that of late-type galaxies, 
%$z=4$ LBGs are positively and even very strongly biased on small scale. 
%The reason for this difference
%is not clear, but it would indicate that 
%multiple star-forming galaxies are more likely
%hosted by a single dark halo at $z=4$ than $z=0$,
%thus suggesting that galaxy formation efficiency 
%(i.e. star-formation rates per mass) 
%may be very high in dark halos at high redshifts.
%
%% revision other3 end
%

\begin{figure}
\epsscale{1.20}
%\plotone{image/acorr_BRiLBG_cumulative.eps}
\plotone{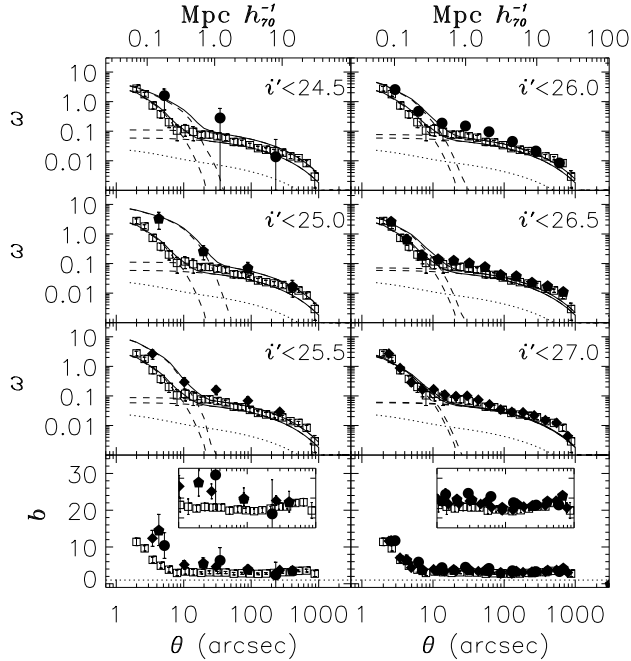}
\caption{
The ACFs of magnitude-limited subsamples 
of LBGs at $z=4.0$. In the top to third-top panels,
the filled symbols are the ACFs of our LBGs
with the limiting magnitude indicated in the legend.
Each of these panels shows the ACF
of $i'<27.5$ LBGs with open squares.
The dotted curves are the ACF of 
dark matter predicted by the non-linear model
of \citet{peacock1996}.
The thick solid and dashed lines indicate the best-fit ACFs of 
the halo model and the breakdown of 1-halo and 2-halo terms for each
subsample, while the thin lines are for $i'<27.5$ LBGs.
In the bottom panels, biases of LBGs for the each magnitude-limited subsample
are presented with the symbols which correspond to
those marks found in the top to third-top panels. 
The plots of large-scale biases are magnified in the inserted boxes.
\label{fig:acorr_BRiLBG_cumulative}}
\end{figure}

%\clearpage

\begin{figure}
%\epsscale{0.8}
\epsscale{1.20}
%\plotone{image/mag_bias_beta_z4.eps}
\plotone{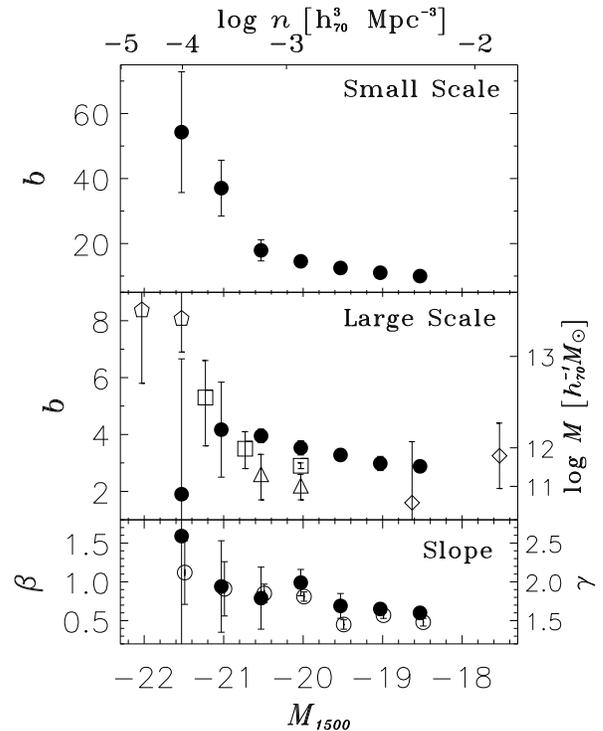}
\caption{
Bias and slope of $z=4$ LBGs
as a function of limiting-absolute magnitude 
calculated from $i'-M_{1500}=46.0$.
Top and middle panels present bias
of small-scale ($2''<\theta<3''$) 
and large-scale ($40''<\theta<400''$) 
clustering. 
Filled circles plot for our LBGs, whose
bias is directly measured from the ACFs.
Open pentagons, squares, triangles, and diamonds present 
the large-scale ($\simeq 8 h_{100}^{-1}$) bias 
estimated from the conventional power-law fit
by \citet{allen2005,ouchi2004b,ouchi2001,arnouts2002}.
In the middle panel, the right-hand vertical scale means
mass of dark halos corresponding to the linear bias \citep{sheth1999}.
The upper abscissa axis ticks number densities of our LBGs.
Bottom panel shows the slope of a power law for the ACFs.
%Filled circles denote the slope for
%large-scale clustering ($40''<\theta<400''$; $\beta_{\rm L}$) 
%with no IC correction.
%Open circles indicate the slope for the all scales ($2''<\theta<1000''$)
%with IC correction, which correspond to results of the
%conventional power-law approximation presented in Table \ref{tab:results}.
Filled and open circles indicate the slopes for large-scale 
($40''<\theta<400''$) with no IC correction ($\beta_{\rm L}$),
and for all scales ($2''<\theta<1000''$) with IC correction ($\beta$; see Table 1),
respectively.
\label{fig:mag_bias_z4}}
\end{figure}

%\clearpage

%\subsection{Luminosity Dependence of Small and Large Scale Clustering}
\subsection{Luminosity Dependence of Clustering}
\label{subsec:luminosity_dependence}
We calculate $\omega(\theta)$ and $b$ for six 
subsamples with limiting magnitudes of 
$i'<24.5$, $25.0$, $25.5$, $26.0$, $26.5$, and $27.0$
(Figure \ref{fig:acorr_BRiLBG_cumulative}).
%
%The small- ($2''<\theta<3''$; $0.05-0.07 h_{100}^{-1}$ Mpc) 
%and large- ($40''< \theta < 400''$; $1-10 h_{100}^{-1}$ Mpc) scale 
%biases for each subsample
%are shown in Figure \ref{fig:mag_bias_z4}.
%
%
%% revision other4 start
%
%{\bf
We define the large- and small-scale biases 
as the biases in the range $40''< \theta < 400''$ 
($1-10 h_{100}^{-1}$ Mpc) and 
$2''<\theta<3''$ ($0.05-0.07 h_{100}^{-1}$ Mpc), respectively,
and show the biases in Figure \ref{fig:mag_bias_z4}.
Although the angular range for the small-scale bias is 
somewhat arbitrary,
the range defined here is beyond the internal structures of
galaxies and below the radii of dark halos with $10^{11} M_\odot$,
and thus is sensitive to a multiplicity of LBGs in a halo.
%Figure \ref{fig:mag_bias_z4} shows the small- and large-scale biases.
%}
%
%% revision other4 end
%
Luminosity segregation of large-scale clustering 
is reported for $z=4$ LBGs 
\citep{ouchi2004b,allen2005,lee2005}.
In Figures \ref{fig:acorr_BRiLBG_cumulative} and 
\ref{fig:mag_bias_z4}, we find 
that ACFs and biases monotonically decrease
from $i'<24.5$ to $27.5$ on small scale
as well as large scale.
Interestingly, Figure \ref{fig:mag_bias_z4} shows that
the small-scale bias has a stronger dependence 
on luminosity ($b\simeq 10-50$) than the large-scale
bias ($b\simeq 3-4$),
and the bottom panels of Figure \ref{fig:acorr_BRiLBG_cumulative}
indicate that
outskirts of small-scale excess
extend to $\theta\sim 10''$ for bright ($i'<24.5-25.5$) LBGs.
All the features of luminosity dependence
suggest that
bright LBGs reside in more massive dark halos, 
since massive dark halos
have not only a high large-scale bias, but also 
a high small-scale bias 
(i.e. high probability of pair galaxies in a massive halo)
and an extended outskirt of bias due to a large halo size.

Although the ACFs depart from a power law,
we approximate the ACFs with a power law, 
in order to compare our results with previous results.
We fit the ACF over $2-1000''$ with 
$\omega(\theta)=A_\omega (\theta^{-\beta} - IC/A_\omega )$,
where IC is the integral constraint \citep{groth1977}.
The bottom panel of Figure \ref{fig:mag_bias_z4} presents
the best-fit slopes, $\beta$, as a function of magnitude. The slopes,
$\beta$, become flatter at faint magnitudes 
(see also \citealt{kashikawa2005}).
This luminosity dependence of $\beta$ is explained by the strong luminosity 
dependence of small-scale clustering as discussed above.
%
%Then we calculate the Limber integral equation 
%and estimate the correlation lengths, $r_0$, 
%which is the normalization of the spatial 
%two-point correlation function, $\xi=(r/r_0)^{-\gamma}$,
%where $\gamma=\beta+1$
%(see \citealt{ouchi2004b} for these calculations).
%We apply the redshift distribution function 
%used in \citet{ouchi2004b}.
%
%% revision cut start
%
Then we calculate the Limber equation with
the redshift distribution function of \citet{ouchi2004b},
and estimate the correlation lengths, $r_0$, 
of spatial two-point correlation function, $\xi=(r/r_0)^{-\gamma}$,
where $\gamma=\beta+1$.
%
%% revision cut end
%
Table \ref{tab:results}
presents the best-fit parameters, $A_\omega$ and $\beta$,
together with $r_0$. These $r_0$ are consistent with
those obtained by 
\citet{ouchi2004b} as well as
by \citet{hildebrandt2004} 
and 
\citet{kashikawa2005}.
However, our results are not consistent with those of
small-sky surveys in HDF,
if we assume that ACF does not significantly evolve
%between $z=3$ and 4 \citep{ouchi2004b}. 
%
%% revision cut start
%
between $z=3$ and 4.
%
%% revision cut end
%
For examples,
\citet{giavalisco2001}
find a small correlation length of 
$r_0=1.0^{+0.8}_{-0.7} h_{100}^{-1}$ Mpc
for $z=3$ LBGs,
while we find a
larger value, $r_0=3.8^{+0.2}_{-0.2} h_{100}^{-1}$ Mpc, 
for our $i'<27.5$ LBGs whose number density is
comparable to that of \citet{giavalisco2001}.
%($1\times10^{-2}$ Mpc$^{-3}$).
%($1\times10^{-2} h_{70}^{3}$ Mpc$^{-3}$). 
We restrict our power-law fitting
to the same narrow range as \citet{giavalisco2001} 
($1''\lesssim \theta \lesssim 20''$), and then 
%we obtain the consistent results within a $2\sigma$ level, i.e.
%
%% revision cut start
%
we obtain the consistent results within errors, i.e.
%
%% revision cut end
%
%$r_0=1.3\pm0.3 h_{100}^{-1}$ Mpc and $\beta=1.9$, 
$r_0=1.3\pm0.3 h_{100}^{-1}$ Mpc and $\beta=1.9\pm0.2$, 
due to the fitting only
to the small-scale excess of the ACF (see \citealt{kravtsov2004}).
Similarly, the correlation length of red galaxies in HDF-S
($r_0 = 8 h_{100}^{-1}$ Mpc; \citealt{daddi2003})
is probably overestimated by the extrapolation from a bump of
small-scale ACF with a relatively flat slope of $\beta=0.8$,
which is also claimed by the model of \citet{zheng2004}.

\subsection{Comparison with a Halo Model}
\label{subsec:comparison}
We fit the halo model of \citet{hamana2004}
to the observed $\omega(\theta)$ and number density, $n$,
simultaneously.
This model predicts the $\omega(\theta)$ and $n$ of galaxies 
contributed by a combination of the 1-halo and 2-halo terms 
in the framework of the CDM model.
The best-fit models are shown 
in Figure \ref{fig:acorr_BRiLBG_cumulative} 
and Table \ref{tab:results}.
These models account for the overall shape of our ACFs,
i.e., the small-scale excess as well as the large-scale clustering
(see also \citealt{lee2005}),
although there remain large residuals
\footnote{
Note that bootstrap errors of the ACF are assumed to be
independent.
% of angular separations.
} 
(e.g. $\chi^2/{\rm dof}=3.0$ for $i'<27.5$ LBGs).
Reducing these residuals results in a decrease in 
the combined likelihood ($\omega(\theta)$ $+$ $n$) from 
the best-fit value. 
%
%The main reason of the residuals is that the model predicts
%a smaller amplitude of $\omega(\theta)$ at a given $n$.
%%
The large residuals imply that
we need a more precise model for our LBGs.
For implications of the model fitting,
Table \ref{tab:results} summarizes
the average number of LBGs in a halo, $\left<N_g\right>$, and
the average masses of the halo, $\left<M_h\right>$, 
\citep{hamana2004} for the best-fit models.
%}
%
The average mass of hosting halos monotonically decreases from
$2\times 10^{12} h_{70}^{-1}M_\odot$ ($i'<24.5$)
to $6\times 10^{11} h_{70}^{-1}M_\odot$ ($i'<27.5$).
The average number of LBGs in a halo
is less than unity, $\left <N_g\right> \simeq 0.2-0.7$,
while the model ACF in Figure \ref{fig:acorr_BRiLBG_cumulative} 
shows a significant 1-halo term produced by multiple LBGs
in one halo.
This implies that majority of halos with an average mass have no LBG
and only some halos host one or multiple LBG(s).
%
%% revision cut start
%
%A more detailed discussion about model fitting will be
%presented in our forthcoming paper.
%
%% revision cut end
%

%% Included in this acknowledgments section are examples of the
%% AASTeX hypertext markup commands. Use \url without the optional [HREF]
%% argument when you want to print the url directly in the text. Otherwise,
%% use either \url or \anchor, with the HREF as the first argument and the
%% text to be printed in the second.

\acknowledgments
We thank M. Fall, M. Giavalisco, K. Lee, S. Okamura, and Z. Zheng
for helpful comments and discussion.

\begin{deluxetable}{crccccccc}
\tablecolumns{11}
\tabletypesize{\scriptsize}
\tablecaption{Summary of Clustering Properties
\label{tab:results}}
\tablewidth{0pt}
\tablehead{
\colhead{} & \colhead{} & \colhead{} & 
\multicolumn{3}{c}{Conventional Power-Law Approx.} &
\colhead{} & 
\multicolumn{2}{c}{Model$\tablenotemark{e}$} \\
\cline{4-6} \cline{8-9} \\
\colhead{$i'_{\rm AB}$} &
\colhead{$N(<i')$\tablenotemark{a}} &
\colhead{$n(<i')$\tablenotemark{b}} &
\colhead{$A_\omega$\tablenotemark{c}} &
\colhead{$\beta$\tablenotemark{c}} &
\colhead{$r_0$\tablenotemark{c}} &
\colhead{$\beta_{\rm L}$\tablenotemark{d}} &
\colhead{$\left<Ng\right>$} &
\colhead{$\log \left<M_{\rm h}\right>$} \\
\colhead{(mag)} & \colhead{} & 
\colhead{($h_{70}^{3}$ Mpc$^{-3}$)} &
\colhead{(arcsec$^{\beta}$)} & \colhead{} &
\colhead{($h_{100}^{-1}$ Mpc)} &
\colhead{} & 
\colhead{} & \colhead{($h_{70}^{-1}$ $M_\odot$)} 
}
\startdata
24.5 & 239 & $9.8 \pm 1.6 \times 10^{-5}$ & $10.5 \pm 8.2$ & $1.1 \pm 0.4$ & $4.9^{+4.3}_{-4.1}$ & $\simeq 1.6$ & $0.2^{+0.2}_{-0.2}$ & $12.3^{+0.1}_{-0.6}$ \\ 
25.0 & 808 & $2.8 \pm 0.3 \times 10^{-4}$ & $5.0 \pm 9.1$ & $0.9 \pm 0.3$ & $5.5^{+1.7}_{-2.1}$ & $0.9 \pm 0.6$ & $0.3^{+0.4}_{-0.3}$ & $12.3^{+0.1}_{-0.2}$ \\
25.5 & 2231 & $6.4 \pm 0.6 \times 10^{-4}$ & $3.1 \pm 1.6$ & $0.8 \pm 0.1$ & $5.0^{+0.7}_{-0.8}$ & $0.8 \pm 0.4$ & $0.6^{+0.1}_{-0.5}$ & $12.1^{+0.1}_{-0.1}$ \\
26.0 & 4891 & $1.3 \pm 0.1 \times 10^{-3}$ & $2.6 \pm 0.6$ & $0.8 \pm 0.1$ & $5.0^{+0.4}_{-0.4}$ & $1.0 \pm 0.2$ & $0.6^{+0.1}_{-0.1}$ & $12.0^{+0.1}_{-0.1}$ \\
26.5 & 8639 & $2.2 \pm 0.3 \times 10^{-3}$ & $0.6 \pm 0.1$ & $0.5 \pm 0.1$ & $4.8^{+0.2}_{-0.3}$ & $0.7 \pm 0.2$ & $0.6^{+0.1}_{-0.1}$ & $11.9^{+0.05}_{-0.05}$ \\
27.0 & 12921 & $3.7 \pm 0.7 \times 10^{-3}$ & $0.8 \pm 0.1$ & $0.6 \pm 0.1$ & $4.4^{+0.1}_{-0.2}$ & $0.7 \pm 0.1$ & $0.6^{+0.1}_{-0.2}$ & $11.8^{+0.07}_{-0.04}$ \\
27.5 & 16920 & $5.8 \pm 1.4 \times 10^{-3}$ & $0.5 \pm 0.1$ & $0.5 \pm 0.1$ & $3.8^{+0.2}_{-0.2}$ & $0.6 \pm 0.1$ & $0.7^{+0.2}_{-0.1}$ & $11.8^{+0.02}_{-0.05}$ \\
\enddata
%\tablecomments{
%}
\tablenotetext{a}{Cumulative numbers. 
Differential surface densities are 
$0.002 \pm 0.001$,
$0.014 \pm 0.002$,
$0.049 \pm 0.004$,
$0.158 \pm 0.007$,
$0.395 \pm 0.011$,
$0.739 \pm 0.014$,
$1.041 \pm 0.017$,
$1.189 \pm 0.018$, and
$1.111 \pm 0.018$
%$0.724 \pm 0.014$ 
arcmin$^{-2}$ (0.5mag)$^{-1}$ for
$i'=$ 23.25,
23.75,
24.25,
24.75,
25.25,
25.75,
26.25,
26.75, and
27.25,
%27.75, 
respectively, which are consistent 
with previous measurements (e.g. \citealt{ouchi2004a}).
}
\tablenotetext{b}{Cumulative number density 
calculated from luminosity function
of \citet{giavalisco2005}.
% who update the work of \citealt{ouchi2004a}.
}
\tablenotetext{c}{Results from the conventional power-law approximation, i.e.
$\omega(\theta)=A_\omega(\theta^{-\beta}-IC/A_\omega)$, 
over $2''-1000''$.
For integral constraints, IC, we apply
$IC/A_\omega= [3,14,21,28,364,154,293] \times 10^{-4}$
for $i'=[24.5,25.0,25.5,26.0,26.5,27.0,27.5]$.
}
\tablenotetext{d}{Power-law slope for the fit of 
$\omega=A_{\omega{\rm L}} \theta^{-\beta_{\rm L}}$ over $40''-400''$
%which corresponds
%to $1-10 h_{100}^{-1}$ Mpc.
with no IC correction.
}
\tablenotetext{e}{
$\chi^2/{\rm dof}=[0.7,0.4,2.5,6.9,8.3,7.1,3.0]$ 
for $i'=[24.5,25.0,25.5,26.0,26.5,27.0,27.5]$.
}
\end{deluxetable}

\end{document}